\documentclass[a4paper]{article}
\usepackage{graphicx}
\usepackage{amsmath}

\newtheorem{theorem}{Theorem}

\newtheorem{definition}[theorem]{Definition}

\newtheorem{remark}[theorem]{Remark}

\input{tcilatex}

\begin{document}

\author{S. Manoff \\
\textit{Bulgarian Academy of Sciences}\\
\textit{Institute for Nuclear Research and Nuclear Energy}\\
\textit{Department of Theoretical Physics}\\
\textit{Blvd. Tzarigradsko Chaussee 72}\\
\textit{1784 Sofia - Bulgaria}}
\title{Frames of reference in spaces with affine connections and metrics}
\date{\textit{e-mail address: smanov@inrne.bas.bg} }
\maketitle

\begin{abstract}
A generalized definition of a frame of reference in spaces with affine
connections and metrics is proposed based on the set of the following
differential-geometric objects: (a) a non-null (non-isotropic) vector field,
(b) the orthogonal to the vector field sub space, (c) an affine connection
and the related to it covariant differential operator determining a
transport along the given non-null vector filed. On the grounds of this
definition other definitions related to the notions of accelerated,
inertial, proper accelerated and proper inertial frames of reference are
introduced and applied to some mathematical models for the space-time. The
auto-parallel equation is obtained as an Euler-Lagrange's equation.
Einstein's theory of gravitation appears as a theory for determination of a
special frame of reference (with the gravitational force as inertial force)
by means of the metrics and the characteristics of a material distribution.

\end{abstract}

\section{Introduction.}

\subsection{Definitions of a frame of reference in classical and
relativistic physics}

The notion of frame of reference is one of the important notions in physics.
It is related, from the one hand, to the mathematical models for description
of physical systems and the space-time, and, on the other hand, to the
experimental check-up of these models. From philosophical and physical point
of view there are many problems in finding out an appropriate definition for
inertial and non-inertial (accelerated) frames of reference in classical
mechanics \cite{Linez} and in relativistic physics \cite{Roditchev}, \cite
{Roditchev-1}, \cite{Mizkjewitsch}, \cite{Vladimirov}, \cite{Ivanitskaja-1}, 
\cite{Ivanitskaja-2}.

In classical mechanics limited notions of absolute rigid body and of
material point are used for solving theoretical and experimental problems in
many cases without any difficulties \cite{Linez}.

From mathematical point of view the application of limited motions is
related to mathematical objects (such as points, curves, co-ordinates etc.)
used in definitions of a frame of reference ($FR$). Unfortunately, until now
there is no general acceptable definition of a $FR$ as well as of
transitions from one to another $FR$.

There are at least three types of methods for defining a $FR$. They are
based on three different basic assumptions:

(a) \textit{Co-ordinate's methods}. A frame of reference is identified with
a local (or global) chart (co-ordinates) in the differentiable manifold $M$ (%
$\dim M=n$, $n\geq 2$) considered as a model of space-time \cite{Nesterov}.
Additional conditions for the transformations of the co-ordinates are also
proposed to ensure the transition from one to another $FR$ [see for
instance, \cite{Roditchev-1} and the references there]. After Roditchev \cite
{Roditchev-1} these groups of transformations of the co-ordinates are called 
\textit{transformations of type} $A$.

(b) \textit{Tetrad's methods}. A frame of reference is identified with a set
of basic contravariant vector fields [$n$ linear independent vectors (called 
$n$-Beins, $n$-beams) $\{e_\alpha \}\in T_x(M)$, $\alpha =1,...,n$, at every
given point $x$ of the manifold $M$ considered as a model of space-time].
For $n=4$, four linear independent contravariant vectors (or their
components in a given basis) are called a tetrad. Additional conditions are
imposed for determining a special type of a $n$-Bein used as a $FR$ \cite
{Mizkjewitsch}, \cite{Roditchev-1}, \cite{Vladimirov}, \cite{Ivanitskaja-1}, 
\cite{Ivanitskaja-2}, \cite{Moeller}. The transition from one to another $FR$
is related to the transformation properties of the $n$-Beins. The groups of
transformations of the $n$-Beins is called group of \textit{transformations
of type} $B$.

(c) \textit{Monad's methods}. A frame of reference is identified with a
non-null (non-isotropic) (time-like) contravariant vector field interpreted
as the velocity of an observer (material point). A transition from one to
another $FR$ is related to the transformations of the contravariant vector
field to another vector field of the same type. Some additional conditions
are required for finding out the explicit form of these transformations
related to the notions of inertial and non-inertial (accelerated) $FR$.
These groups of transformations of the type $u^{\prime }=\Omega (u)$, $%
\Omega \in \otimes ^1\,_1(M)$, $u^{\prime }$, $u\in T(M)$, are considered by
Roditchev \cite{Roditchev}, \cite{Roditchev-1}, where $\Omega $ is called
affinor. Transformations of this type could be called \textit{%
transformations of type} $C$.

If a contravariant vector field at a point of the manifold $M$ is defined as
a vector belonging to an introduced at this point $n$-Bein and further the
vectors of the $n$-Bein are considered as tangential vectors to the
co-ordinates given in a neighbourhood of this point, then all methods for
determining a $FR$ could have physical interpretation. Usually, the main
requirements to all physical laws are of two types:

(a) All physical laws should be expressed analytically in a general
covariant form with respect to transformations of the types $A$ and $B$,
i.e. they should be represented by means of tensors general covariant with
respect to $A$ and $B$ \cite{Roditchev}.

(b) All physical laws should be general covariant with respect to
transformations of types $A$ and $B$.

The cited above methods are used for describing physical interactions \cite
{Gogsadze} and especially the gravitational interaction \cite{Ivanitskaja-1}%
, \cite{Ivanitskaja-2}, \cite{Hehl}.

In this paper a generalization of the notion of $FR$ is proposed and
considered on the basis of an ''extended'' monad formalism. In Section 2 the
notions of different types of frames of reference (inertial, accelerated,
proper inertial, proper accelerated etc.) are introduced. In Section 3
transitions from one to another $FR$ are determined on the basis of well
established external covariant differential operators. In Section 4 some
invariant properties of parallel and auto-parallel equations are recalled
for the case of spaces of affine connections and metrics [$(\overline{L}%
_n,g) $, $(L_n,g)$-spaces, and their special cases]. The auto-parallel
equation is obtained as Euler-Lagrange's equation.

\begin{remark}
The reader is kindly asked to refer to \cite{Manoff-3}, \cite{Manoff-4}, or 
\cite{Manoff-5}, where all basic symbols and definitions used without
explanations in this paper are introduced.
\end{remark}

\subsection{What are $(\overline{L}_n,g)$-spaces?}

The main characteristics of a $(\overline{L}_{n},g)$-space which differ from
these of a $(L_{n},g)$-space are based on a weaker definition of the notion
of dual bases in finite dual vector spaces over a differentiable manifold $M$
\cite{Manoff-5}, \cite{Manoff-6}. Instead of the common in multilinear
algebra (canonical) contraction operator $C$ acting on basic vectors $%
\{e^{\alpha }(x)\}\in N_{1\mid \,x\,\in \,M}$ and on basic vectors $%
\{e_{\beta }(x)\}\in N_{2\mid \,x\,\in \,M}$ of the vector spaces $N_{1}$
and $N_{2}$ with equal (finite) dimensions $(dim\,N_{1}=dim\,N_{2}=n)$ at a
point $x\in M$: 
\begin{eqnarray*}
C &:&\left( e^{\alpha }(x)\text{, }e_{\beta }(x)\right) \rightarrow C\left(
e^{\alpha }(x)\text{, }e_{\beta }(x)\right) \equiv e^{\alpha }(e_{\beta
})_{\mid \,x\,\in \,M}=\delta _{\beta }^{\alpha }\equiv g_{\beta }^{\alpha }
\\
\text{with }\delta _{\beta }^{\alpha } &\equiv &g_{\beta }^{\alpha }=1\text{
for }\alpha =\beta \text{ and }\delta _{\beta }^{\alpha }\equiv g_{\beta
}^{\alpha }=0\text{ for }\alpha \neq \beta \text{ ,}
\end{eqnarray*}

\noindent a contraction operator $S$ has been used: 
\begin{eqnarray*}
S &:&\left( e^{\alpha }(x)\text{, }e_{\beta }(x)\right) \rightarrow S\left(
e^{\alpha }(x)\text{, }e_{\beta }(x)\right) \equiv e^{\alpha }(e_{\beta
})_{\mid \,x\,\in \,M}=f^{\alpha }\,_{\beta }(x) \\
\,\,\text{with \ \ \ \thinspace }det(f^{\alpha }\,_{\beta }) &\neq &0
\end{eqnarray*}

Thus, \textit{the definition of dual vector bases has been weaken for vector
spaces} (as fibres of vector bundles) \textit{over a differentiable manifold 
}$M$. [The weaken definition of dual vector bases is meaningless in the
multilinear algebra but it is very interesting in the case of vector spaces
over a differentiable manifold.] It leads to the possibility of introducing
two different affine connections (whose components differ not only by sign)
for the tangent and cotangent vector spaces and, respectively, to two
different affine connections for contravariant and covariant tensor fields
over $M$. All formulas written in index-free form are identical and valid in
their form (but not in their contents) for $(L_n,g)$- and $(\overline{L}%
_n,g) $-spaces. The difference between them appear only if they are written
in a given (co-ordinate or non-co-ordinate) basis. For instance, 
\begin{equation*}
g(u,u)=g_{ij}\cdot u^i\cdot u^j
\end{equation*}

\noindent in a $(L_n,g)$-space, but 
\begin{eqnarray*}
g(u,u) &=&g_{ij}\cdot f^i\,_k\cdot f^j\,_l\cdot u^k\cdot u^l=g_{ij}\cdot u^{%
\overline{i}}\cdot u^{\overline{j}}=g_{\overline{k}\overline{l}}\cdot
u^k\cdot u^l \\
\text{with }f^i\,_j(x^l) &=&S(dx^i\text{,\thinspace }\partial _j)=S(\partial
_j\text{,\thinspace }dx^i)=dx^i(\partial _j)
\end{eqnarray*}

\noindent in a $(\overline{L}_n,g)$-space; 
\begin{equation*}
g(u)=g_{kl}\cdot u^l\cdot dx^k
\end{equation*}

\noindent in a $(L_n,g)$-space, but 
\begin{equation*}
g(u)=g_{kl}\cdot f^l\,_m\cdot u^m\cdot dx^k=g_{kl}\cdot u^{\overline{l}%
}\cdot dx^k=g_{k\overline{m}}\cdot u^m\cdot dx^k
\end{equation*}

\noindent in a $(\overline{L}_n,g)$-space. In a $(L_n,g)$-space 
\begin{equation*}
S(dx^i\text{,\thinspace }\partial _j)=C(dx^i\text{,\thinspace }\partial
_j)=\delta _j^i\equiv g_j^i\text{ .}
\end{equation*}

All formulas with indices can be very easily specialized for $(L_n,g)$%
-spaces by omitting the bars of all indices and taking into account that $%
\delta _{j;k}^i\equiv g_{j;k}^i=0$. $\delta _j^i$ are the \textit{Kronecker
symbols} identical with the components $g_j^i$ of the \textit{Kronecker
tensor} $Kr=g_j^i\cdot \partial _i\otimes dx^j=g_\beta ^\alpha \cdot
e_\alpha \otimes e^\beta $.

\section{Generalized definition of a frame of reference}

Let us take a closer look at the third type of methods for introduction of a
frame of reference. The method is called by different names [method of $\tau 
$-field \cite{Mizkjewitsch}, monad formalism, $[(n-1)+1]$-representation,
conception of the solitary (lonely) observer etc.]. It is proposed by Eckard
(1941) \cite{Eckart} and Lief (1951) [see \cite{Mizkjewitsch}] and later on
refound and applied by many authors \cite{Uhlmann}, \cite{Schmutzer}. The
consideration of a vector field $u$ as the velocity of an observer induces
(at least locally) a tangential sub space $T^{\perp u}(M)$, orthogonal to $u$
over $M$. All observed physical events and systems are projected to the
direction of the vector field $u$ and its sub space by the use of the
corresponding to $u$ contravariant and covariant metrics $h^{u}=\overline{g}-%
\frac{1}{e}\cdot u\otimes u$ and $h_{u}=g-\frac{1}{e}\cdot g(u)\otimes g(u)$%
. The invariant projections of the different tensor characteristics of an
observed physical system were then given the corresponding physical
interpretation \cite{Schmutzer}, \cite{Manoff-1}. The world line of the
observer is determined by its tangential vector $u$ at every of its points.
A tangential vector $u$ at a point $x\in M$, considered as an initial point,
is transported to all other points of the world line by means of a transport 
\cite{Manoff-3}, \cite{Manoff-4} determined by a preliminary given
contravariant affine (linear) connection. With other words, the world line
of the observer is determined by the transport of its tangential vector and,
therefore, by the affine connection respectively. On this basis, we can
conclude that \textit{it} \textit{is not enough for a definition of a frame
of reference a non-null (time-like) vector field }$u$\textit{\ [with its
corresponding orthogonal to it sub space }$T^{\perp u}(M)$\textit{] to be
given}. A $FR$ should be determined in the third type of methods (c) by the
set of four geometric objects:

(a) A non-null [time-like (if $\dim M=4$)] vector field $u\in T(M)$.

(b) Orthogonal to $u$ tangential sub space $T_x^{\perp u}(M)$ at every point 
$x\in M$, where $u$ is defined.

(c) (Contravariant) affine connection $\nabla =\Gamma $. It determines the
type of transport along the trajectory to which $u$ is a tangent vector
field. $\Gamma $ is related to the covariant differential operator $\nabla
_{u}$ along $u$ \cite{Manoff-3}, \cite{Manoff-5}.

(d) Metric tensor field $g$ at every point $x\in M$, where $u$ is defined.
It enables one to measure lenghts and angles.

\begin{remark}
In the further considerations the last point (d) will not be considered
explicitly. It is only assumed that a covariant metric tensor field $g$ and
its corresponding contravariant vector field $\overline{g}$ exist among the
other structures of a frame of reference. The existence of the metric tensor
field $g$ is implicitly given in the definition of $T^{\perp u}(M)$.
\end{remark}

Now we can define the notion of frame of reference.

\begin{definition}
The set $FR\sim \lbrack u$, $T^{\perp u}(M)$, $\nabla =\Gamma $, $\nabla
_{u}]$ is called frame of reference in a differentiable manifold $M$
considered as a model of the space or of the space-time.
\end{definition}

Every $FR$ determines a congruence of curves (world lines) (a set of
non-intersecting curves) in the range of the definition of the vector field $%
u$ and the action of the covariant differential operator $\nabla _u$.

A differentiable manifold $M$, provided with affine connection $\nabla $,
i.e. the pair $(M,\nabla )$, is called space with affine connection \cite
{Manoff-3} - \cite{Manoff-6}. A space with affine connection and metrics
determines the existence of a set of frames of reference $FR\sim \lbrack u$, 
$T^{\perp u}(M)$, $\nabla =\Gamma $, $\nabla _{u}]$ which is invariant with
respect to transformations of the types $A$ and $B$. The different types of
frames of reference [inertial, non-inertial (accelerated) frame of
reference] could be defined on the grounds of the above basic definition.
For every special type of a $FR$ additional conditions should be imposed
characterizing its properties related to the action of $\nabla _{u}$ on the
vector field $u$ and on its corresponding orthogonal vector fields $\xi
_{\perp }\in T^{\perp u}(M)$.

Since $\nabla =\Gamma $ could fulfil the condition $\nabla =\Gamma =0$ under
a special choice of the basic contravariant vector fields \cite{Iliev} - 
\cite{Iliev-6}, \cite{Hartley} at least at a point or on a trajectory to
which $u$ is a tangent vector field, it is always possible a $FRIF$ to be
transform to an $IFR$ (frame of reference without inertial forces).

The existence of a $FR$ with inertial forces is due to the fact that in some
cases inappropriate basic vector fields $\{e_\alpha \}$ are chosen. In the
table below $^F\nabla _u$ denotes a Fermi-Walker transport \cite{Manoff-3}, 
\cite{Manoff-4}.

\vspace{10 pt}%
%
{}

\textbf{Table 1}.\textit{\ Different types of frames of reference}

\begin{center}
\begin{tabular}{lll}
Type of a frame of reference & Symbol & Conditions, determining \\ 
&  & the type of a frame of reference \\ 
\_\_\_\_\_\_\_\_\_\_\_\_\_\_\_\_\_\_ & \_\_\_\_ & \_\_\_\_\_\_\_\_\_\_\_\_\_%
\_\_\_\_\_\_ \\ 
General frame of reference & $FR$ & $[u,\,T^{\perp u}(M),\,\nabla =\Gamma $, 
$\nabla _u]$ \\ 
&  &  \\ 
Accelerated frame of reference & $AFR$ & $FR+[\,\nabla _uu=a\neq 0]$ \\ 
Inertial frame of reference & $IFR$ & $FR+[\,\nabla =\Gamma =0$, $\,\,\nabla
_uu=a=0]$ \\ 
Proper accelerated & $PAFR$ & $FR+[\,\,\nabla _uu=a\neq 0,^F\nabla _u=\nabla
_u-\overline{A}_u$, \\ 
frame of reference &  & \thinspace $^F\nabla _u\xi _{\perp }=0$ , $\xi
_{\perp }\in T^{\perp u}(M)]$ \\ 
Proper inertial & $PIFR$ & $FR+[\nabla =\Gamma =0$, $\,\,\nabla _uu=a=0$ ,
\\ 
frame of reference &  & $^F\nabla _u=\nabla _u-\overline{A}_u$ , \\ 
&  & \thinspace $^F\nabla _u\xi _{\perp }=0$ , $\xi _{\perp }\in T^{\perp
u}(M)]$ \\ 
Frame of reference & $IFR$ & $IFR$ \\ 
without inertial forces & $FRWIF$ & $FR+[\nabla =\Gamma =0$, $\,\,\nabla
_uu=a\neq 0]$ \\ 
Frame of reference & $FRIF$ & $FR+[\nabla =\Gamma \neq 0$, $\,\,\nabla
_uu=a=0]$ \\ 
with inertial forces &  & 
\end{tabular}
\end{center}

\section{Transition from one to another frame of reference}

A $FR\sim [u$, $T^{\perp u}(M)$, $\nabla =\Gamma $, $\nabla _u]$ will
determine by means of its affine connection the model of a space or of a
space-time, where the physical systems and events (at least locally) could
be considered. Therefore, a transition from one to another affine connection
could be related to the transition from one to another $FR$. At that, there
are two possibilities for a transition from one to another affine connection.

(a) Transition from$\nabla $ to $\nabla ^{\prime }$ on the grounds of the
transformation properties of $\nabla $ under changing the basic vector
fields (or co-ordinates) used in the corresponding frame of reference. Let
us recall some well known facts.

If the basic vectors $\{\partial _i\}$ or $\{e_i\}$ are transforming as 
\begin{equation*}
e_\alpha =A_\alpha \,^i\cdot \partial _i=A_\alpha \,^k\cdot e_k\text{ ,
\thinspace \thinspace and if\thinspace \thinspace \thinspace \thinspace
\thinspace }\nabla _{e_\beta }e_\alpha =\Gamma _{\alpha \beta }^\gamma \cdot
e_\gamma \text{ , \thinspace }\nabla _{\partial _j}\partial _i=\Gamma
_{ij}^k\cdot \partial _k\text{ ,}
\end{equation*}

\noindent then $\nabla =\Gamma $ will transform in $\nabla ^{\prime }=\Gamma
^{\prime }$ as 
\begin{equation}
\Gamma _{\alpha \beta }^\gamma =A_\alpha \,^i\cdot A_\beta \,^j\cdot
A_k\,^\gamma \cdot \Gamma _{ij}^k+A_\beta \,^j\cdot A_k\,^\gamma \cdot
A_\alpha \,^k\,_{,j}\text{ .}  \label{2.1}
\end{equation}

A transformation of the above type (as a transformation of type $A$ or $B$)
does not change the form of the transports of the vector field $u$, i.e. $%
\nabla _uu=a$ is invariant to the transition of $\nabla =\Gamma $ to $\nabla
^{\prime }=\Gamma ^{\prime }$. To every affine connection $\nabla =\Gamma $
and a vector field $u$ corresponds a covariant differential operator $\nabla
_u$ invariant under the above transformation of the basic vector fields
(co-ordinates).

(b) Transition from $\nabla =\Gamma $ to $^e\nabla =\overline{\Gamma }$ on
the grounds of a change (deformation) of $\Gamma $ by the use of a tensor
field of third rank $\overline{A}:=\overline{A}^i\,_{jk}\cdot \partial
_i\otimes dx^j\otimes dx^k\in \otimes ^1\,_2(M)$ in the form \cite{Manoff-3}%
, \cite{Manoff-4} 
\begin{equation}
\overline{\Gamma }\,_{jk}^i=\Gamma _{jk}^i-\overline{A}\,_{jk}^i\text{ .}
\label{2.2}
\end{equation}

The corresponding to $\overline{\Gamma }$ covariant differential operator $%
^e\nabla _u$ (called \textit{extended covariant differential operator})
could be written in the form 
\begin{equation}
^e\nabla _u=\nabla _u-\overline{A}_u\text{ ,}  \label{2.3}
\end{equation}

\noindent where $\overline{A}_u:=\overline{A}\,^i\,_{j\overline{k}}\cdot
u^k\cdot \partial _i\otimes dx^j$.

This type of transformation is invariant under the change of the basic
vector fields (and co-ordinates). It induces a new frame of reference $[u$, $%
T^{\perp u}(M)$, $^e\nabla $, $^e\nabla _uu]$ which being induced by the
frame of reference $[u$, $T^{\perp u}(M)$, $\nabla $, $\nabla _uu]$ differs
from it. Therefore, we can relate a transition from one to another $FR$ to a
transition from one $\nabla _u$ to another $^e\nabla _u$ covariant
differential operator acting on tensor fields over a differentiable manifold 
$M$.

A general transition from one to another $FR$ could have the form 
\begin{equation}
Transition:[u,\,\,T^{\perp u}(M),\,\,\,\nabla =\Gamma ,\,\,\nabla
_u]\longrightarrow [\overline{u},T^{\perp \overline{u}}(M),\,\,^e\nabla =%
\overline{\Gamma },\,\,^e\nabla _{\overline{u}}]\text{ ,}  \label{2.4}
\end{equation}

\noindent where 
\begin{equation}
\begin{array}{c}
\overline{u}:=\Omega (u)=\overline{u}^i\cdot \partial _i=\Omega ^i\,_{%
\overline{j}}\cdot u^j\cdot \partial _i\text{ ,\thinspace \thinspace
\thinspace \thinspace \thinspace \thinspace \thinspace \thinspace }\Omega
\in \otimes ^1\,_1(M)\text{ ,} \\ 
T^{\perp \overline{u}}=\{\overline{\xi }\in T(M):g(\overline{u},\overline{%
\xi })=0\}\text{ , \thinspace \thinspace \thinspace \thinspace \thinspace
\thinspace }^e\nabla =\nabla -\overline{A}\text{ .}
\end{array}
\label{2.5}
\end{equation}

The last type of transformations of $\Gamma \,$, $D:\Gamma \rightarrow 
\overline{\Gamma }$, could be called \textit{transformations of type} $D$
(because they are related to deformations of an affine connection $\Gamma $).

\subsection{Transition from an accelerated frame of reference to an inertial
frame of reference}

A transition from an $AFR\sim $ $[u,\,\,T^{\perp u}(M),\,\,\,\nabla
,\,\,\nabla _uu=a\neq 0]$ to an $IFR\sim [u,\,\,T^{\perp
u}(M),\,\,^e\,\nabla =0$,$\,\,^e\nabla _uu=0]$ could be interpreted as a
transition from an accelerated $FR$ to an inertial $FR$ under conserving the
vector field (velocity) of the observer. This means that the vector field in
both frames of reference is one and the same but the transport of this
vector field is different in the two frames of reference. This type of
transition could be done by two steps.

(a) Transition from $[u$,$\,\,T^{\perp u}(M)$,$\,\,\,\nabla =\Gamma $, $%
\nabla _{u}$,$\,\,\nabla _{u}u=a\neq 0]$ to

$[u$,$\,\,T^{\perp u}(M)$,$\,\,^{e}\nabla =\overline{\Gamma }\neq 0$, $%
^{e}\nabla _{u}$, $\,^{e}\,\nabla _{u}u=0]$, i.e. $AFR\longrightarrow FRIF$.

(b) Transition from $[u$,$\,\,T^{\perp u}(M)$,$\,\,^{e}\,\nabla =\overline{%
\Gamma }\neq 0$,$\,\,^{e}\nabla _{u}$, $\,^{e}\,\nabla _{u}u=0]$ to

$[u$,$\,\,T^{\perp u}(M)$,$\,\,^{e}\,\nabla =\overline{\Gamma }=0$, $%
^{e}\nabla _{u}$, $\,^{e}\,\nabla _{u}u=0]$, i.e. $FRIF\longrightarrow IFR$.

Let us considered every step separately from the other.

\subsubsection{Transition from an accelerated frame of reference to a frame
of reference with inertial forces}

The finding out of a transition from $[u$,$\,\,T^{\perp u}(M)$,$\,\,\,\nabla
=\Gamma $, $\nabla _u$,$\,\,\nabla _uu=a\neq 0]$ to $[u$,$\,\,T^{\perp u}(M)$%
,$\,\,^e\,\nabla =\overline{\Gamma }\neq 0$, $^e\nabla _u$, $\,^e\,\nabla
_uu=0]$ is related to the problem of finding out an extended covariant
differential operator $^e\nabla _u=\nabla _u-\overline{A}_u$ such that $%
^e\nabla _uu=0$.

The equation $\nabla _uu=0$ could be written in a co-ordinate basis in a $(%
\overline{L}_n,g)$-space \cite{Manoff-2} in the form 
\begin{equation}
u^i\,_{;j}\cdot u^j=u^i\,_{,j}\cdot u^j+\Gamma _{jk}^i\cdot u^j\cdot u^k=a^i%
\text{ ,}  \label{2.6}
\end{equation}

\noindent or in the form 
\begin{equation}
u^i\,_{,j}\cdot u^j+(\Gamma _{jk}^i-\frac 1e\cdot a^i\cdot g_{\overline{k}%
\overline{j}})\cdot u^k\cdot u^j=0\text{ , \thinspace \thinspace \thinspace
\thinspace \thinspace \thinspace \thinspace \thinspace \thinspace \thinspace
\thinspace }e=g(u,u)\neq 0\text{ .}  \label{2.7}
\end{equation}

On the other side, $^e\nabla _uu=0$ could be written in the form 
\begin{equation*}
u^i\,_{:j}\cdot u^j=u^i\,_{,j}\cdot u^j+\overline{\Gamma }\,_{jk}^i\cdot
u^j\cdot u^k=u^i\,_{,j}\cdot u^j+(\Gamma _{jk}^i-\overline{A}%
\,^i\,_{jk})\cdot u^j\cdot u^k=0\text{ .}
\end{equation*}

If we chose the form of $\overline{A}\,^i\,_{jk}$ as $\overline{A}%
\,^i\,_{jk}=\frac 1e\cdot a^i\cdot g_{\overline{j}\overline{k}}$, then we
find $\overline{\Gamma }\,^i\,_{jk}$ in the form 
\begin{equation}  \label{2.8}
\overline{\Gamma }\,_{jk}^i=\Gamma _{jk}^i-\frac 1e\cdot a^i\cdot g_{%
\overline{j}\overline{k}}\text{ }.
\end{equation}

The tensor $\overline{A}_u$ has the form 
\begin{equation}  \label{2.9}
\overline{A}_u=\frac 1e\cdot a\otimes g(u)=\frac 1e\cdot a^i\cdot g_{j%
\overline{k}}\cdot u^k\cdot \partial _i\otimes dx^j\text{ .}
\end{equation}

Therefore, $^e\nabla _uu=\nabla _uu-a=0$.

In the new frame of reference $[u$, $T^{\perp u}(M)$, $^e\nabla =\overline{%
\Gamma }$, $^e\nabla _u$, $^e\nabla _uu=0]$ the vector field appears as an
auto-parallel vector field with respect to the new affine connection $%
^e\nabla =\overline{\Gamma }$. On this basis we can make the conclusion that 
\textit{every ''real'' force inducing an acceleration }$a$\textit{\ with
respect to an affine connection }$\nabla =\Gamma $\textit{\ could be
considered as an inertial force with respect to the corresponding to }$%
\nabla =\Gamma $\textit{\ affine connection }$^e\nabla =\overline{\Gamma }$%
\textit{\ fulfilling the condition} (\ref{2.8}).

The action of the extended contravariant differential operator $^e\nabla _u$
on a contravariant vector field $\xi \in T(M)$ can be found as 
\begin{equation}  \label{2.10}
^e\nabla _u\xi =\nabla _u\xi -\frac 1e\cdot a\cdot g(u,\xi )=\nabla _u\xi
-\frac le\cdot a\text{ , \thinspace \thinspace \thinspace \thinspace
\thinspace \thinspace }l=g(u,\xi )\text{ .}
\end{equation}

If $\xi $ is orthogonal to $u$, i.e. if $l=0$, then $^e\nabla _u\xi _{\perp
}=\nabla _u\xi _{\perp }$, $\xi _{\perp }=\overline{g}[h_u(\xi )]$.
Therefore, the new extended covariant differential operator $^e\nabla _u$
does not change the type of the transport (induced by $\nabla _u$) of the
vectors of the sub space $T^{\perp u}(M)$.

\subsubsection{Transition from a frame of reference with inertial forces to
an inertial frame of reference}

Let us now consider the transition from $[u$, $T^{\perp u}(M)$, $\nabla
=\Gamma \neq 0$, $\nabla _u$, $\nabla _uu=0]$ to $[u$, $T^{\perp u}(M)$, $%
\nabla =\Gamma =0$, $\nabla _u$, $\nabla _uu=0]$. From 
\begin{equation*}
\begin{array}{c}
\nabla _{e_j}e_i=\Gamma _{ij}^k\cdot e_k\text{ , \thinspace \thinspace
\thinspace \thinspace }e_\alpha =A_\alpha \,^i\cdot e_i\text{ , \thinspace
\thinspace \thinspace \thinspace }e_i=A_i\,^\beta \cdot e_\beta \text{ ,
\thinspace \thinspace \thinspace }A_\alpha \,^i\neq 0\text{ ,} \\ 
A_\alpha \,^i\cdot A_i\,^\beta =g_\alpha ^\beta \text{ , \thinspace
\thinspace \thinspace \thinspace \thinspace \thinspace }A_i\,^\beta \cdot
A_\beta \,^j=g_i^j\text{ \thinspace , }
\end{array}
\end{equation*}

\noindent and from the condition 
\begin{equation}
\nabla _{e_\beta }e_\alpha =\overline{\Gamma }\,_{\alpha \beta }^\gamma
\cdot e_\gamma =0\text{ ,}  \label{2.12}
\end{equation}

\noindent we get 
\begin{equation}
A_\alpha \,^k\,_{,j}+A_\alpha \,^i\cdot \Gamma _{ij}^k=0\text{ .}
\label{2.13}
\end{equation}

The Latin indices $i$, $j$, $k$, $...$, belong to the indices of the basis $%
\{e_k$, $k=1,...,n\}$ and the Greek index $\alpha $ belongs to the indices
of the basis $\{e_\alpha \}$. Therefore, for every given index $\alpha $ and 
$j$, and given components $\Gamma _{ij}^k$ of the affine connection $\Gamma $
in the basis $\{e_k\}$ we have a system of $k$ partial differential
equations for $A_\alpha \,^i$ in the type (\ref{2.13}). Further, if we
consider the basis $\{e_k\}$ as a co-ordinate basis, i.e. $%
\{e_k\}:=\{\partial _k\}$, then we can write the system of partial
differential equations (PDEs) in the same form, but $A_\alpha \,^k$ could be
considered as components $A^k$ of $\alpha $ vector fields $A_\alpha
=A_\alpha \,^k\cdot \partial _k$. Thus, the system of PDEs appears as $%
\alpha $ equations for the components of $\alpha $ vector fields $A_\alpha $
in a co-ordinate basis which are transported parallel to a basic vector
field $\partial _j$ with respect to the affine connection $\Gamma $ with
components $\Gamma _{ij}^k$ in a co-ordinate basis $\{\partial _k\}$: 
\begin{equation}
\nabla _{\partial _j}A_\alpha =0\,\,\,\,\,\,\,\,\cong
\,\,\,\,\,\,\,\,A_\alpha \,^k\,_{;j}=0\text{ .}  \label{2.14}
\end{equation}

The last equations mean that we have to find $\alpha $ covariant constant
vector fields $A_\alpha $ with respect to the affine connection $\nabla
=\Gamma $.

The system of PDEs (\ref{2.14}) appears as a necessary and sufficient
condition for $\overline{\Gamma }=0$ (with components $\overline{\Gamma }%
_{\alpha \beta }^\gamma =0$ in the contravariant vector basis $\{e_\alpha \}$%
). The proof of this statement is trivial if we use the above relations.

A necessary but not sufficient condition following from (\ref{2.13}) is the
condition for a given vector field $u$: 
\begin{equation}  \label{2.15}
\nabla _uA_\alpha =0\text{ , \thinspace \thinspace \thinspace \thinspace
\thinspace \thinspace }u=\frac d{ds}=\frac{dx^i}{ds}\cdot \partial _i\in T(M)
\end{equation}

The last $\alpha $ equations are equivalent to the equations for $\alpha $
vector fields $A$ transported parallel along an auto-parallel vector field $%
u $ ($\nabla _uu=0$) 
\begin{equation}
\frac{dA_\alpha \,^k}{ds}+\Gamma _{ij}^k\cdot A_\alpha \,^i\cdot \frac{dx^j}{%
ds}=0\text{ , \thinspace \thinspace \thinspace \thinspace \thinspace
\thinspace \thinspace \thinspace }\alpha =1,\,...\,,\,n\text{ .}
\label{2.16}
\end{equation}

Since the vector field $u=u^k\cdot \partial _k$ is transported parallel to
itself, we can find $n-1$ additional vector fields $A_\alpha $ (with $\alpha
=1$, $...$, $n-1$) transported parallel to $u$ and orthogonal to $u$. The
solutions of the equations (\ref{2.16}) determine $\alpha $ vector fields $%
A_\alpha (s)$ transported parallel along a curve with parameter $s$ and
tangential to it vector field $u$. The components $A_\alpha {}^k(s)$ of
these $\alpha $ vector fields $A_\alpha (s)$ in the basis $\{e_k=\partial
_k\}$ determine the components $A_\alpha \,^k(s)$ of the transformation
matrix $\left( A_\alpha \,^k(s)\right) $ for a transition from the basis $%
\{e_k\}$ to the basis $\{e_\alpha \}$ along with the transition from $\Gamma
(s)$ with components in $\{e_k=\partial _k\}$ to $\overline{\Gamma }(s)=0$
with vanishing components $\overline{\Gamma }_{\alpha \beta }^\gamma (s)=0$
in the new basis $\{e_\alpha \}$.

\begin{remark}
The above considerations are a short representation of the results given in 
\cite{Iliev} - \cite{Iliev-6} and \cite{Hartley} expressed in a more evident
for our investigations form.
\end{remark}

\subsubsection{Relations between the components of the (contravariant)
curvature tensor for the affine connection $\Gamma $ and that for the affine
connection $\overline{\Gamma }$}

The components $R^i\,_{jkl}$ of the (contravariant) curvature tensor $Riem$
induced by the affine connection $\Gamma $ could be written in a co-ordinate
basis $\{\partial _k\}$ in the form 
\begin{equation*}
R^i\,_{jkl}=\Gamma _{jl,k}^i-\Gamma _{jk,l}^i+\Gamma _{jl}^m\cdot \Gamma
_{mk}^i-\Gamma _{jk}^m\cdot \Gamma _{ml}^i\text{ .}
\end{equation*}

The components $\overline{R}\,_{jkl}^i$ of the (contravariant) curvature
tensor $\overline{Riem}$ induced by the affine connection $\overline{\Gamma }
$ written in the same co-ordinate basis will have the form 
\begin{equation*}
\overline{R}^i\,_{jkl}=\overline{\Gamma }\,_{jl,k}^i-\overline{\Gamma }%
\,_{jk,l}^i+\overline{\Gamma }\,_{jl}^m\cdot \overline{\Gamma }\,_{mk}^i-%
\overline{\Gamma }\,_{jk}^m\cdot \overline{\Gamma }\,_{ml}^i\text{ .}
\end{equation*}

By the use of (\ref{2.8}) we can express $\overline{R}^i\,_{jkl}$ by means
of $R^i\,_{jkl}$ in the form 
\begin{equation*}
\overline{R}^i\,_{jkl}=R^i\,_{jkl}+F^i\,_{jkl}\text{ ,}
\end{equation*}

\noindent where 
\begin{eqnarray}
F^i\,_{jkl} &=&(\frac 1e\cdot a^i\cdot g_{\overline{j}\overline{k}%
})_{,l}-(\frac 1e\cdot a^i\cdot g_{\overline{j}\overline{l}})_{,k}+  \notag
\\
&&+\frac 1{e^2}\cdot a^i\cdot a^m\cdot (g_{\overline{j}\overline{l}}\cdot g_{%
\overline{m}\overline{k}}-g_{\overline{j}\overline{k}}\cdot g_{\overline{m}%
\overline{l}})+  \notag \\
&&+\frac 1e\cdot (a^m\cdot g_{\overline{j}\overline{k}}\cdot \Gamma
_{ml}^i+a^i\cdot g_{\overline{m}\overline{l}}\cdot \Gamma _{jk}^m)+  \notag
\\
&&+\frac 1e\cdot (a^m\cdot g_{\overline{j}\overline{l}}\cdot \Gamma
_{mk}^i+a^i\cdot g_{\overline{m}\overline{k}}\cdot \Gamma _{jl}^m)\text{ .}
\label{2.16a}
\end{eqnarray}

\textit{Special case}: Flat space $M_n$: $\Gamma _{jk}^i\equiv
0:R_{jkl}^i\equiv 0$. 
\begin{eqnarray}
\overline{R}^i\,_{jkl} &=&F^i\,_{jkl}=(\frac 1e\cdot a^i\cdot g_{\overline{j}%
\overline{k}})_{,l}-(\frac 1e\cdot a^i\cdot g_{\overline{j}\overline{l}%
})_{,k}+  \label{2.16b} \\
&&+\frac 1{e^2}\cdot a^i\cdot a^m\cdot (g_{\overline{j}\overline{l}}\cdot g_{%
\overline{m}\overline{k}}-g_{\overline{j}\overline{k}}\cdot g_{\overline{m}%
\overline{l}})\text{ .}  \notag
\end{eqnarray}

The curvature components $\overline{R}^i\,_{jkl}$ of the corresponding
space-time are induced by the acceleration $a=a^i\cdot \partial _i$ of the
observer (related to its accelerated frame of reference). On the other hand, 
\begin{equation}
a^i=\frac en\cdot (\Gamma _{jk}^i-\overline{\Gamma }\,_{jk}^i)\cdot g^{jk}%
\text{ , \thinspace \thinspace \thinspace \thinspace }n=4\,\,\,\text{%
,\thinspace \thinspace \thinspace \thinspace }e=g(u,u)\neq 0\text{ .}
\label{2.16c}
\end{equation}

Every acceleration $a$ could be determined by the difference of two affine
connections, the existing metrics, and the contravariant vector field $u$.
For a flat space ($\Gamma _{jk}^i=0$), $a^i=-\frac en\cdot \overline{\Gamma }%
\,_{jk}^i\cdot g^{jk}$. Therefore, every field theory in a flat space-time
(corresponding to an accelerated $FR$ without inertial forces) using the
equation $\nabla _uu=a$ as an equation for a particle moving with an
acceleration $a$ in an external field could be considered as a field theory
in a $(\overline{L}_n,g)$-space, where the same particle moves without
acceleration, i. e. its velocity obeys an auto-parallel equation ($^e\nabla
_uu=0$) in a $(\overline{L}_n,g)$-space (corresponding to a $FR$ with
inertial forces).

\subsection{Transition preserving the form of an auto-parallel equation}

Let us now consider the transition of a $FR$ of the type $[u$, $T^{\perp
u}(M)$, $\nabla $, $\nabla _uu=0]$ to a $FR\sim $ $[u$, $T^{\perp u}(M)$, $%
^e\nabla $, $^e\nabla _uu=0]$. The following problem could be investigated
and some solutions of it could be found.

\textit{Problem}. Let an auto-parallel equation $\nabla _uu=0$ be given with
respect to a contravariant affine connection $\nabla =\Gamma $. Find a new
contravariant affine connection $^e\nabla =\overline{\Gamma }$ with a
corresponding extended covariant differential operator $^e\nabla _u$ such
that $^e\nabla _uu=0$.

\textit{Solution}. The auto-parallel equation $\nabla _uu=0$ (with respect
to $\nabla $) and the new auto-parallel equation $^e\nabla _uu=0$ (with
respect to $^e\nabla $) can be written in the forms respectively 
\begin{equation}
u^i\,_{,j}\cdot u^j+\Gamma _{jk}^i\cdot u^j\cdot u^k=0\text{ , \thinspace
\thinspace \thinspace \thinspace \thinspace \thinspace \thinspace \thinspace 
}u^i\,_{,j}+(\Gamma _{jk}^i-\overline{A}\,^i\,_{\overline{j}\overline{k}%
})\cdot u^j\cdot u^k=0\text{ .}  \label{2.17}
\end{equation}

After substraction of the second equation from the first one, we obtain a
condition for $\overline{A}\,^i\,_{\overline{j}\overline{k}}\cdot u^j\cdot
u^k$ in the form 
\begin{equation}  \label{2.18}
\overline{A}\,^i\,_{\overline{j}\overline{k}}\cdot u^j\cdot u^k=0\text{ .}
\end{equation}

Different solutions are possible for $\overline{A}\,^i\,_{\overline{j}%
\overline{k}}$ fulfilling the above condition. Let us give some of them.

1. $\overline{A}\,^i\,_{\overline{j}\overline{k}}:=q^i\cdot h_{\overline{j}%
\overline{k}}$. The tensor $\overline{A}$ will have the form 
\begin{equation}  \label{2.19}
\overline{A}=q\otimes h_u=q^i\cdot h_{jk}\cdot \partial _i\otimes
dx^j.dx^k\in \otimes ^1\,_2(M)\,\,\,\,\text{, \thinspace \thinspace
\thinspace \thinspace }q\in T(M)\text{ ,}
\end{equation}

\noindent and the condition $\overline{A}_u=\overline{A}(u)=0$ is fulfilled
leading to the equation $^e\nabla _uu=$ $\nabla _uu=0$.

2. $\overline{A}\,^i\,_{\overline{j}\overline{k}}:=t^i\cdot \omega _{%
\overline{j}\overline{k}}+q^i\cdot h_{\overline{j}\overline{k}}$. The tensor 
$\overline{A}$ will have the form 
\begin{equation}  \label{2.20}
\overline{A}=t\otimes \omega +q\otimes h_u\text{ , \thinspace \thinspace
\thinspace }\omega \in \Lambda ^2(M)\text{ , \thinspace \thinspace }t,q\in
T(M)\text{ ,}
\end{equation}

\noindent and the condition $\overline{A}_u(u)=\overline{A}(u,u)=0$ is
fulfilled.

3. $\overline{A}\,^i\,_{\overline{j}\overline{k}}:=t^i\cdot \omega _{%
\overline{j}\overline{k}}+q^i\cdot (h_{\overline{j}\overline{l}}\cdot \eta
^l\cdot p_k+h_{\overline{k}\overline{l}}\cdot \eta ^l\cdot p_j)$. The tensor 
$\overline{A}$ will have the form 
\begin{equation}
\overline{A}=t\otimes \omega +q\otimes [h_u(\eta )\otimes p+p\otimes
h_u(\eta )]\text{ ,\thinspace \thinspace \thinspace \thinspace \thinspace
\thinspace \thinspace }t,q,\eta \in T(M)\text{ ,\thinspace \thinspace
\thinspace \thinspace \thinspace \thinspace }p\in T^{*}(M)\text{ ,}
\label{2.21}
\end{equation}

\noindent and the condition $\overline{A}_u(u)=\overline{A}(u,u)=0$ is also
fulfilled.

All three possible solutions for $\overline{A}\,^i\,_{\overline{j}\overline{k%
}}$ and the corresponding solutions for $\overline{A}$ preserve the form of $%
\nabla _uu=0$. On the other hand, solutions 2. and 3. for $\overline{A}$ and
respectively for $^e\nabla _u$ induce new type of transports (different from
that induced by $\nabla _u$) for the vector fields orthogonal to $u$:

\begin{equation}
(a)\,\,\,\,\,\,\,\,^e\nabla _u\xi =\nabla _u\xi -\overline{A}_u(\xi )=\nabla
_u\xi -\omega (u,\xi )\cdot t\,\,,\,\,\,\,\,\,\xi ,\,\,t,\,\,u\in T(M).
\label{2.22}
\end{equation}

\begin{equation}
(b)\,\,\,\,\,\,\,\,\,\,\,\,\,^e\nabla _u\xi =\nabla _u\xi -\overline{A}%
_u(\xi )=\nabla _u\xi -\omega (u,\xi )\cdot t-p(u)\cdot h_u(\eta ,\xi )\cdot
q.  \label{2.23}
\end{equation}

Therefore, the transitions of the type $\nabla _u\rightarrow \,^e\nabla _u$,
leading to the forminvariance of $\nabla _uu=0$, do not lead in general to
the invariance of the transport of the vector fields orthogonal to $u$.

\section{Canonical representation of the parallel and the auto-parallel
equations}

\subsection{Canonical representation of a parallel equation}

Let a congruence $x^i(\tau ,\lambda )$ be given described by the two
parameters $\tau $ and $\lambda $ and by the tangent vector fields 
\begin{equation*}
u:=\frac \partial {\partial \tau }=\frac{\partial x^i}{\partial \tau }\cdot
\partial _i\text{ , \thinspace \thinspace \thinspace \thinspace \thinspace
\thinspace \thinspace \thinspace \thinspace \thinspace and \thinspace
\thinspace \thinspace \thinspace \thinspace \thinspace \thinspace \thinspace
\thinspace }\xi :=\frac \partial {\partial \lambda }=\frac{\partial x^j}{%
\partial \lambda }\cdot \partial _j\text{ }
\end{equation*}

\noindent respectively. Let us consider a parallel transport of the vector
field $\xi $ along the vector field $u$%
\begin{equation}
\nabla _u\xi =f\cdot \xi \text{ , \thinspace \thinspace \thinspace
\thinspace \thinspace \thinspace \thinspace \thinspace }f\in C^r(M)\text{ .}
\label{3.1}
\end{equation}

An equation of this type is called recurrent equation (or recurrent relation
for the vector field $\xi $) \cite{Sinjukov}. Three types of invariance of
this equation could be found.

(a) Invariance with respect to a change of the co-ordinates (charts) or the
basic vector fields in the manifold $M$. This invariance (of type $A$ or $B$%
) is obvious because it follows from the index-free form of the equation.

(b) Forminvariance with respect to a change of the vector $\xi $ with a
collinear to it vector $\eta :=\varphi \cdot \xi $.

\begin{remark}
In $(\overline{L}_{n},g)$-spaces the transformation $\xi \rightarrow \eta $
can not be related to a transformation of type $C$ $[\Omega :\xi \rightarrow
\Omega (\xi ):=\eta :=\varphi \cdot \xi $, with $\Omega =\varphi \cdot Kr]$
as it is the case in $(L_{n},g)$-spaces.
\end{remark}

(c) Forminvariance with respect to a change of the parameter $\lambda $
determining $\xi $.

The proofs of all types of invariance are trivial.

\subsection{Canonical representation of an auto-parallel equation}

Let us consider the auto-parallel equation $\nabla _uu=f\cdot u$ as a
special case of a parallel equation for $\xi =u$, $f=f(x^k(\tau ))$. In this
case 
\begin{equation*}
\lambda =\tau \text{ \thinspace \thinspace \thinspace ,\thinspace \thinspace
\thinspace \thinspace }u=\frac d{d\tau }\text{ \thinspace \thinspace
\thinspace \thinspace , \thinspace \thinspace \thinspace \thinspace }u^i=%
\frac{dx^i}{d\tau }\text{ \thinspace \thinspace \thinspace and \thinspace
\thinspace \thinspace \thinspace \thinspace }\sigma =\sigma (\tau )\text{ ,
\thinspace \thinspace \thinspace }\tau =\tau (\sigma )\text{ , }\frac{%
d\sigma }{d\tau }\neq 0\text{ .}
\end{equation*}

Then 
\begin{equation*}
u^i=\frac{dx^i}{d\sigma }\cdot \frac{d\sigma }{d\tau }=\overline{u}^i\cdot 
\frac{d\sigma }{d\tau }\text{ , \thinspace \thinspace \thinspace \thinspace
\thinspace }\overline{u}^i=\frac{dx^i}{d\sigma }\text{ ,}
\end{equation*}
\begin{equation*}
u^i\,_{,j}\cdot u^j=\frac{d\overline{u}^i}{d\sigma }\cdot \left( \frac{%
d\sigma }{d\tau }\right) ^2+\overline{u}^i\cdot \frac{d^2\sigma }{d\tau ^2}%
\text{ ,}
\end{equation*}
\begin{equation*}
u^i\,_{;j}\cdot u^j-f\cdot u^i=\left( \frac{d\sigma }{d\tau }\right) ^2\cdot
\left( \frac{d\overline{u}^i}{d\sigma }+\Gamma _{jk}^i\cdot \overline{u}%
^j\cdot u^k\right) +\overline{u}^i\cdot \left( \frac{d^2\sigma }{d\tau ^2}%
-f(\tau )\cdot \frac{d\sigma }{d\tau }\right) =0\text{ .}
\end{equation*}

One can chose as condition for determining the function $\sigma =\sigma
(\tau )$ as a function of $\tau $ the vanishing of the last term of the
above equation 
\begin{equation*}
\frac{d^2\sigma }{d\tau ^2}-f(\tau )\cdot \frac{d\sigma }{d\tau }=0\text{ .}
\end{equation*}

The last equation is of the type 
\begin{equation*}
y^{\prime }-f\cdot y=0\text{ , \thinspace \thinspace where\thinspace
\thinspace }\,\,\,\,\,\,\,\,\,y=\frac{d\sigma }{d\tau }\text{ ,}%
\,\,\,\,\,\,\,\,\,\,y^{\prime }=\frac{d^2\sigma .}{d\tau ^2}\text{
.\thinspace \thinspace }
\end{equation*}

Then 
\begin{equation*}
\sigma =\sigma _0+\sigma _1\cdot \int \exp \left( \int f(\tau )\cdot d\tau
\right) \text{ , \thinspace }\sigma _0=\text{ const., }\sigma _1=\text{
const.}
\end{equation*}

After the introduction of the new parameter $\sigma =\sigma (\tau )$ (called
canonical parameter), the auto-parallel equation will have the form 
\begin{equation*}
\nabla _{\overline{u}}\overline{u}=0\text{ \thinspace \thinspace \thinspace }%
\cong \,\,\,\,\overline{u}^i\,_{;j}\cdot u^j=0\,\,\,\,\,\text{\thinspace ,
\thinspace \thinspace \thinspace \thinspace \thinspace \thinspace \thinspace
\thinspace \thinspace }\overline{u}=\frac d{d\sigma }\text{ , }\overline{u}%
^i=\frac{dx^i}{d\sigma }\text{ .}
\end{equation*}

The last equation for $\overline{u}$ is called auto-parallel equation in
canonical form.

\subsection{Auto-parallel equations as Euler-Lagrange's equations}

In pseudo-Riemannian spaces without torsion ($V_n$-spaces) the geodesic
equation (identical with the auto-parallel equation $\nabla _uu=0$) can be
found on the ground of the variation $\delta S=0$ of an action $S$
identified with the length of a curve with parameter $s$%
\begin{equation*}
S=\int ds+s_0\text{ }:\delta S=0\Rightarrow \nabla _uu=0\text{ , with }%
\nabla =\Gamma =\{\,\,\,\}\text{ .}
\end{equation*}

[$\{\,\}$ is the Levi Civita (symmetric) affine connection.] The same method
has been used for finding out the geodesic equation in a $(\overline{L}_n,g)$%
-space \cite{Manoff-2}. Since the geodesic equation (interpreted as an
equation for motion of a moving free test particle in an external
gravitational field) differs from the auto-parallel equation in a $(%
\overline{L}_n,g)$- or $(L_n,g)$-space, the old question arises as what is
the right equation for description of a free moving particle in a $(%
\overline{L}_n,g)$-space: the geodesic equation (G) or the auto-parallel
equation (A). The most authors \textit{believe} that the geodesic equation
is the more appropriate equation. One of their major arguments is that the
geodesic equation is related to a variational principle (as a basic
principle in classical physics) in contrast to the auto-parallel equation.
So the so called G-A problem induces investigations of possible ways for
finding its solution \cite{Saa-1}-\cite{Saa-5}, \cite{Kleinert}, \cite
{Fiziev-1}, \cite{Fiziev-2}. Unfortunately, no general solution was found
until now. In our opinion, the failure is related to the attempt of using
analogous variational expression as in the case of the geodesic equation. In 
$(\overline{L}_n,g)$- and $(L_n,g)$-spaces the auto-parallel equation has
much more complicated structure (related to torsion and nonmetricity) than
the geodesic equation. On the other side, an auto-parallel contravariant
vector field induces additional structures such as the orthogonal to it sub
space $T^{\perp u}(M)$ which should be taken into account if we wish to find
the auto-parallel equation on the basis of a variational principle. If we
use the basic arguments for introducing a generalized definition of a $FR$
we can also find a solution of the G-A problem by the use of the method of
Lagrangians with covariant derivatives (MLCD) \cite{Manoff-5}.

Let us define a Lagrangian invariant in the form 
\begin{eqnarray}
L &=&p_0+h_0\cdot g[\nabla _u(\rho \cdot u)\text{,}\xi ]=p_0+h_0\cdot g_{%
\overline{i}\overline{j}}\cdot (\rho u^i)_{;k}\cdot u^k\cdot \xi ^j\text{ ,}
\notag \\
\text{ \thinspace \thinspace \thinspace \thinspace }p_0\text{, }h_0 &=&\text{
const., \thinspace \thinspace }\rho \in C^r(M)\text{ , \thinspace }u\text{,}%
\xi \in T(M)\text{ .}\,\,  \label{3.2}
\end{eqnarray}

\noindent with the additional condition for the contravariant vector fields $%
u$ and $\xi $: $g(u,\xi )=l=0$. The corresponding action $S$ could be
written in the form 
\begin{equation}
S=\int \sqrt{-d_g}\cdot (L+\lambda \cdot l)\cdot d^{(n)}x=\int (L+\lambda
\cdot l)\cdot d\overline{\omega }\text{ , \thinspace \thinspace \thinspace }%
d_g=\det (g_{ij})<0\,\text{\thinspace \thinspace .}  \label{3.3}
\end{equation}

\noindent where $\lambda $ is a Lagrangian multiplier. $L$ is interpreted as
the pressure $p$ of a physical system, $u$ is the velocity of the particles, 
$\rho $ is their proper mass density, and $\xi $ is a vector, orthogonal to $%
u$. By the use of the MLCD we obtain the covariant Euler-Lagrange equations
for the vector fields $u$ and $\xi $ obeying the condition $l=0$%
\begin{equation}
\frac{\delta L}{\delta \xi ^i}=0:\,\,\,\,\,\,\,\,\,\,\,\,\,\,u^i\,_{;j}\cdot
u^j=[\frac \lambda {h_0}-u(\log \rho )]\cdot u^i\text{ ,}  \label{3.4}
\end{equation}
\begin{eqnarray}
\frac{\delta L}{\delta u^i} &=&0:\,\,\,\,\,\xi ^i\,_{;j}\cdot u^j=(q-\delta
u+\frac \lambda {h_0})\cdot \xi ^i+g_{\overline{k}\overline{l}}\cdot
u^k\,_{;j}\cdot g^{ji}\cdot \xi ^l-  \notag \\
&&-[g^{ij}\cdot (g_{\overline{j}\overline{k}})_{;m}\cdot u^m-g_{k;j}^i\cdot
u^j]\cdot \xi ^k\text{ .}  \label{3.5}
\end{eqnarray}
\begin{equation}
\frac{\delta L}{\delta \lambda }=0:g(u,\xi )=l=0\text{ .}  \label{3.5a}
\end{equation}

In index-free form the equations for $u$ and $\xi $ would have the forms: 
\begin{equation}
\nabla _uu=k\cdot u\text{ ,\thinspace \thinspace \thinspace \thinspace
\thinspace \thinspace \thinspace \thinspace \thinspace \thinspace \thinspace
\thinspace \thinspace \thinspace \thinspace \thinspace \thinspace \thinspace
\thinspace \thinspace }k=\frac \lambda {h_0}-u(\log \rho )\text{ ,}
\label{3.6}
\end{equation}
\begin{equation}
\nabla _u\xi =m\cdot \xi +K-N\text{ , \thinspace \thinspace \thinspace
\thinspace \thinspace \thinspace \thinspace \thinspace \thinspace \thinspace 
}m=q-\delta u+\frac \lambda {h_0}\text{ , \thinspace \thinspace \thinspace
\thinspace \thinspace \thinspace \thinspace \thinspace \thinspace }
\label{3.7}
\end{equation}
\begin{equation}
q=q_j\cdot u^j\text{ , \thinspace \thinspace \thinspace \thinspace }%
q_j=T_{kj}\,^k-\frac 12\cdot g^{\overline{k}\overline{l}}\cdot
g_{kl;j}+g_k^l\cdot g_{l;j}^k\text{ ,}  \label{3.8}
\end{equation}
\begin{equation}
\delta u=u^k\,_{;k}\text{ , \thinspace \thinspace \thinspace \thinspace
\thinspace }K=K^i\cdot \partial _i=(g_{\overline{k}\overline{l}}\cdot
u^k\,_{;j}\cdot g^{ji}\cdot \xi ^l)\cdot \partial _i\text{ ,}  \label{3.9}
\end{equation}
\begin{equation}
N=N^i\cdot \partial _i\text{ ,\thinspace \thinspace \thinspace \thinspace
\thinspace \thinspace }N^i=[g^{ij}\cdot (g_{\overline{j}\overline{k}%
})_{;m}\cdot u^m-g_{k;j}^i\cdot u^j]\cdot \xi ^k\text{ .}  \label{3.10}
\end{equation}

The Euler-Lagrange's equation (\ref{3.6}) is just the auto-parallel equation
in a non-canonical form. For $\rho =$ const., it will have the form $\nabla
_uu=\frac \lambda {h_0}\cdot u$. After changing the parameter of the curve
to which $u$ is a tangential vector field the auto-parallel equation could
be found in its canonical form $\nabla _{\overline{u}}\overline{u}=0$.

The Euler-Lagrange's equation for $\xi $ (\ref{3.7}) has in general a more
complicated form than the parallel equation for $\xi $ along $u$ ($\nabla
_u\xi =g\cdot \xi $). For different affine connections (and the
corresponding models of space-time) this equation would have different
solutions. Therefore, if we consider an auto-parallel equation as a result
of a variational principle we should take into account the corresponding
orthogonal to $u$ sub space.

\begin{remark}
The Lagrangian invariant $L$ could be defined without the requirement $\xi $
to be orthogonal to $u$. The covariant Euler-Lagrange's equations will be
then found for $u$ and a vector field $\xi \in T(M)$. For $\rho =$ const.
the auto-parallel equation will have its canonical form. The orthogonality
condition for $u$ and $\xi $ could be introduced after solving the
Euler-Lagrange equations for $u$ or $\xi $.
\end{remark}

\subsection{Einstein's theory of gravitation as a theory for finding out an
appropriate frame of reference for describing the gravitational interaction}

Let us consider the Einstein theory of gravitation (ETG) from the point of
view of the definition of a generalized $FR$. For that we should see how
every element in the definition for a $FR\sim [u$, $T^{\perp u}(M)$, $\nabla
=\Gamma $, $\nabla _u]$ is related to the theory ($\dim M=4$).

The vector field $u$ in the ETG for a material distribution is usually
related to the $4$-velocity of the material points. For Einstein's equations
in vacuum $u$ is not uniquely determined. An assumption is made that a free
particle in an external gravitational field is moving on a geodesic world
line [i.e. its $4$-velocity obeys the auto-parallel equation $\nabla _{u}u=0$
(identical with the geodesic equation in a $V_{4}$-space)]. The affine
connection $\nabla =\Gamma =\{\,\,\}$ (Levi-Civita connection) has
components $\{_{jk}^{i}\}$ in a basis $\{\partial _{k}\}$ determined by the
metric tensor field $g=g_{ij}\cdot dx^{i}.dx^{j}$. These components are
found on the basis of the Einstein equations in vacuum for the metric $g$.
Therefore, ETG from a point of view of a $FR$ is a theory for description of
the gravitational interaction on the basis of an appropriate $FR$ determined
by the use of the Einstein equations, in which $FR$ the gravitational force
in vacuum appears as an inertial force.

\begin{remark}
Conditions under which a sub space $T_{x_{1}}^{\perp u}(M)$ at a given point 
$x_{1}\in M$ on a curve with tangent vector $u$ could (or could not)
intersect the sub space $T_{x_{2}}^{\perp u}(M)$ of another point $x_{2}\in M
$ lying on the same curve as well as Fermi-Walker and conformal transports
related to frames of reference in $(\overline{L}_{n},g)$- and $(L_{n},g)$%
-spaces and different from these already proposed in the literature \cite
{Manoff-3}, \cite{Manoff-4} will be considered elsewhere.
\end{remark}

\section{Conclusion}

In the present paper a generalized definition of the notion of frame of
reference has been introduced and considered. It leads to the hypothesis
that every $FR$ determines a model of space-time used for description of
physical systems and events. For instance, the Einstein theory of
gravitation could be formulated either in a (pseudo) Riemannian space or in
a Minkowski space \cite{Babak}. On the other hand, every type of $FR$ as a
mankind's construction could cause problems in attempts for describing the
physical events in the best possible way \cite{Hayashi}.

$Acknowledgments$

This work is supported by the National Science Foundation of Bulgaria under
Grant No. F-642.

\end{document}